\documentclass{aa}
\usepackage{graphics}

\begin{document}

   \thesaurus{}
   \title{Temporal evolution of the pulse width in GRBs}
   \author{E. Ramirez-Ruiz\inst{1,2} \and E. E. Fenimore\inst{2}}	
   \institute{Facultad de Ciencias, Universidad Nacional Aut\'onoma de
   M\'exico, Distrito Federal, M\'exico 04510\and Los Alamos National
   Laboratory, Mail Stop D436, Los Alamos, NM 87545}
   \maketitle
   \begin{abstract}
   Many cosmological models of GRBs envision the energy source to be a
   cataclysmic stellar event leading to a relativistically expanding
   fireball. Particles are thought
   to be accelerated at shocks and produce nonthermal radiation. The
   highly variable temporal structure observed in most GRBs has
   significantly constrained models. 
   By using different methods of statistical analysis in the time
   domain we show that the width of the pulses in GRBs time histories
   remain remarkably constant throughout the classic GRB phase. If the emission
   sites lie on a relativistically expanding shell, we 
   determine both the amount of    
   deceleration and the angular spread of the emitting region from
   the time dependency of the pulse width. We find no deceleration
   over at least 2/3 of the burst duration and angular spreads of the complete
   emitting shell
   that are substancially smaller than $\Gamma^{-1}$. The lack of temporal
   evolution of the pulse width should be explained by
   any fireball shock scenario. 
   \end{abstract}
	
%________________________________________________________________

\section{Introduction}
 The cosmological origin of GRBs, established as a result of optical
 follow-up observations of fading X-ray counterparts to GRBs, requires
 an extraordinarily large amount of energy to flood the
 entire universe with gamma rays. The lack of apparent
 photon-photon attenuation of high energy photons implies substantial
 bulk relativistic motion. The bulk Lorentz factor,
 $\Gamma=(1-\beta^{2})^{-1/2}$, must be
 on the order of $10^{2}$ to $10^{3}$.
 Two major scenarios involving relativistic shells have been
 developed. In the external shock models, a
 relativistic shell, that expands outward for a long period of time, is
 generated by a single release of energy during
 the merger. The shell coasts in a gamma-ray quiet phase for a certain
 period. Eventually, the shell becomes gamma-ray active due to the
 interactions with the external medium. If the shell
 has a velocity, $v=\beta c$, then the photons emitted $on$ $axis$
 over a period $t'$ (``proper time'' in the comoving frame of the shell)
 arrive at a detector over a much shorter period, $T={t' \over
 2\Gamma}$. The duration of
 the event is set by the expansion of the shell and the complex
 temporal structure is due to inhomogeneities in the shell and/or
 the ambient material.
 The alternative theory is that a central site releases energy in
 the form of a wind or multiple shells over a period of time
 commensurate with the observed duration of GRB. Each subpeak in the
 GRB is the result of a separate explosive event in the central
 site. If the emission sites do indeed lie on a relativistically
 expanding shell, the pulse width in the time GRBs histories scales
 as  $\Delta T=\Lambda\Delta t'$, where $\Lambda$ is the
 Doppler factor, $\Lambda=\Gamma(1-\beta \cos\theta)$. Here
 $\theta$ is the angle of the motion of the emitting region with
 respect to the direction of the emission. In
 this paper, we proposed to determine the angular spread of the emitting
 region and the amount of deceleration from the time histories of
 many GRBs.
\begin{figure}
 \resizebox{\hsize}{!}{\includegraphics{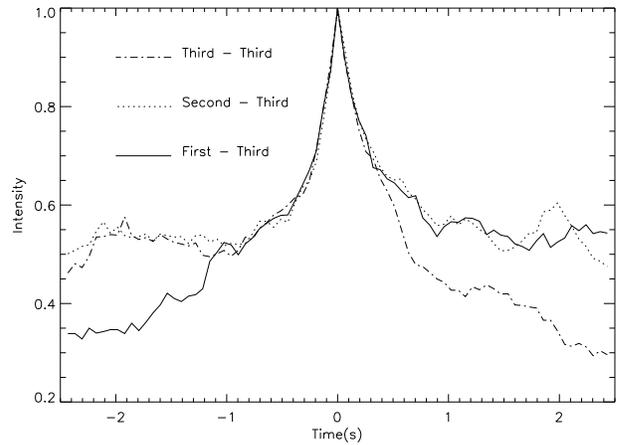}}
 \caption[]{Average peak alignment from 53 bright BATSE bursts with
 durations longer than 20 s.The three curves show the
 average pulse shape for the largest peak in the first third, second
 third, and last third of the bursts. We find no  significant
 change, during the gamma-ray phase, in the average peak width over at
 least 2/3 of $T_{90}$. Models should account this empirical trend
 in GRB physics.} 
\end{figure}

\section{Pulse width evolution obtained from time histories}
%\subsection{The average peak width}
   A visual inspection of the BATSE catalog of multiple-peaked time
   histories reveals that peaks usually have about the same duration at
   the beginning of the burst as near the end of the burst. Our aim is to
   characterize  and measure the pulse shape as a function of arrival time. 
   The aligned peak method measures the average pulse temporal
   structure, each burst contributes to the average by aligning the
   largest peak (\cite{mi97}). We used all 53 bursts from
   the BATSE 4B Catalog that were longer than 20s and brighter than 5
   photons s$^{-1}$ cm$^{-2}$. Each burst must have at least one peak, as
   determined by a peak-finding algorithm (similar to \cite{hli96}), in
   each  third of its duration. The largest peak in each third was
   normalized to unity and
   shifted in time, bringing the largest peaks of all bursts into
   common alignment. This method was applied in each third
   of the duration of the bursts. Thus, we  obtained one curve
   of the averaged pulse shape for each different section of
   the bursts (as shown in Figure 1). The average profile is notably
   identical in each 1/3 of $T_{90}$ (we estimate the differential
   spread, $S$, to be $\sim$ 1\%).
%\subsection{Pulse width from individual bursts}
 
   We have shown the lack of temporal evolution of the peak width in
   the context of an average of many
   bursts. Now, we expand  our analysis to individual bursts. An excellent
   analysis has been provided by \cite{jn96}, where they examined the
   temporal profiles of bright GRBs by fitting those
   profiles with pulses. From the set of bursts that they analysed, we
   used the 28 bursts with the characteristic of having seven or more
   fitted pulses within their duration. To
   obtain the temporal dependency of the pulse width, we selected the
   five largest peaks in each burst and  fit their FWHM to a function
   of the arrival time, $\Delta T=k \left({T-T_{c}\over
   T_{90}} \right )^{\alpha}$.  We chose this model because the
   expected dependency from the ``external'' shock model scales as
   ${T/T_{90} \over \Gamma(1+\beta)}$ (if a relativistic shell is
   responsible for the shape in the time histories, $T_{0}$ should be
   proportional to $T_{90}$; see \cite{fmn96}). The purpose of 
   the $T_{c}$ parameter is to correct the BATSE time to the time
   since the beginning of the explosion. Figure 2 shows the
   distribution of the power law indexes ($\alpha$) for the set of bursts
   analysed. The distribution shows that in individual bursts the
   pulse width does not increase throughout their duration as
   predicted by the external shock model, for which $\alpha$ is expected to be
   1 (or larger if deceleration is ocurring). In fact, most bursts
   show a diminution in pulse width.
\begin{figure}
 \resizebox{\hsize}{!}{\includegraphics{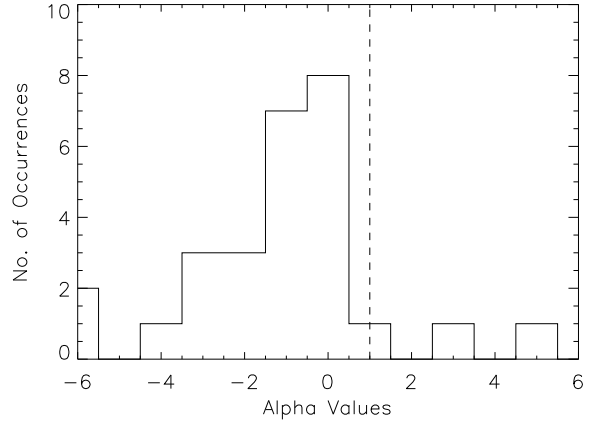}}
 \caption[]{Distribution of the $\alpha$ parameter from 28 bright
 BATSE bursts with durations longer than 1.5 s. The pulse width
 ($\Delta$T), as expected for the external model, should scale as
 $T/T_{90}$. We fit $k \left ({T-T_{c}\over T_{90}}\right)^{\alpha}$
 to find the temporal evolution of the pulse width in each burst. The
 dashed line is the expected index value, $\alpha$, for a single
 relativistic shell ($\alpha$=1). The distribution shows that the vast
 majority of the bursts present no time evolution (or negative) of the
 pulse width.}
\end{figure}
\section{Discussion}
   We have uncovered that the width of the peaks remains remarkably constant
   throughout a burst. The width should scales as
   $\Gamma(1-\beta\cos\theta)$. In the external shock model, such
   lack of temporal evolution implies that $\Gamma$ must be nearly
   constant. Peaks occur late in the burst because
   they are from regions off axis where the delay is caused by the
   curvature of the shell. The later peaks would be wider and delayed
   because off
   axis regions have larger $\theta$'s. A constant peak width indicates
   that all emitting entities must have similar $\theta$. Since the maximum
   angular size of the shell allowed by a  differential spread of $S$
   is $S\Gamma^{-1}$,  the entire size of the shell must be a few
   percent of $\Gamma^{-1}$. Thus, the only external shock model
   that is consistent with the observations is one where the overall
   size of the shell is much smaller than $\Gamma^{-1}$ and 
   there is no deceleration during the classic GRB phase.  This adds
   to our previous arguments (\cite{fmn96}, \cite{frs98}) that a central
   engine (internal shocks) is
   the more likely explanation for the observed choatic 
   time history.  In the context of the internal shock model, we
   cannot place a limit on angular extent but the lack of evolution of
   the peak width indicates that $\Gamma$  must be remarkably
   constant throught the internal shock phase. 

\end{document}